\documentclass[journal]{IEEEtran}
\usepackage{url,amsmath,graphicx}
\usepackage{cite}
\usepackage{graphicx}
\usepackage{psfrag}
\newcommand\Exp{\mathbb{E}}
\usepackage[T1]{fontenc}
\usepackage{selinput}
\SelectInputMappings{%
	adieresis={ä},
	eacute={é},
	Lcaron={Ľ},
}

\usepackage{fancyhdr}
\usepackage{cite}
\usepackage{graphicx}
\usepackage{psfrag}
\usepackage{subfigure}
\usepackage{url}
\usepackage{stfloats}
\usepackage{amsmath}
\usepackage{array}
\usepackage{fancyhdr}
\usepackage{epsfig}
\usepackage{amssymb}
\usepackage{color}

\begin{document}
	\title{ Relay-Assisted Mixed FSO/RF Systems over M\'alaga-$\mathcal{M}$ and $\kappa$-$\mu$ Shadowed Fading Channels}
\author{ Im\`ene~Trigui, Nesrine~Cherif, and Sofi\`ene~Affes\\
INRS-EMT, 800, de la Gaucheti\`{e}re Ouest, Bureau
6900, Montr\'{e}al, H5A 1K6, Qc, Canada. \\
\{itrigui, nesrine.cherif, affes\}@emt.inrs.ca
}
\maketitle

	\begin{abstract}
This letter presents a unified analytical framework for the computation of the ergodic capacity and the outage probability of relay-assisted  mixed FSO/RF transmission.  In addition to accounting for different FSO detection techniques, the mathematical model offers a twofold unification of mixed  FSO/RF systems by considering mixed M\'alaga-$\mathcal{M}$/$\kappa$-$\mu$ shadowed fading, which includes as special cases nearly all linear turbulence/fading models adopted in the open literature.

	\end{abstract}

		%\begin{keywords} Amplify-and-forward (AF), ergodic capacity, free-space optics (FSO), $\kappa$-$\mu$ shadowed fading, M\'alaga-$\mathcal{M}$ distribution, outage probability, pointing errors.\end{keywords}\vspace{-0.3cm}
	\section{Introduction}
	\IEEEPARstart{R}{ecently}, free-space optical (FSO) communications have gained a significant attention due to their advantages of higher bandwidth in unlicensed spectrum and higher throughput compared to their RF counterparts \cite{Yang2}. Hence, the gathering of both FSO and RF technologies arises as a promising solution for securing connectivity between the RF access  and the fiber-optic-based backbone networks. As such, there has been prominent interest in mixed FSO/RF systems where RF transmission is used at one hop and FSO transmission at the other  \cite{yang}-\!\!\cite{zedini}. Most contributions within this  research line  consider restrictive irradiance and channel gain probability density function (PDF) models for the FSO and RF links, respectively. The  most commonly utilized models for the irradiance in FSO links are the lognormal and the Gamma-Gamma ($\mathcal{G}$-$\mathcal{G}$) (\!\!\cite{emmna},\!\cite{zedini} and references therein). Recently, a new generalized statistical model, the M\'alaga-$\mathcal{M}$ distribution, was proposed in \cite{ansari} to model the irradiance fluctuation of an unbounded optical wavefront  propagating through a turbulent medium under all irradiance conditions. Characterized in \cite{navas2} as a mixture of Generalized-${\cal K}$ and  discrete
Binomial distributions, the  Mal\'aga-$\mathcal{M}$ distribution  unifies most statistical models
exploited so far and is able to better reflect a wider range
of turbulence conditions\cite{ansari}, \!\!\cite{navas2}. On the RF side, previous works typically assume either Rayleigh or Nakagami-$m$ fading \cite{emmna},\!\cite{zedini}, thereby lacking the flexibility to account for disparate signal
	propagation mechanisms as those characterized  in 5G communications which will accommodate a wide range of usage scenarios with diverse link requirements. To bridge this gap in the literature, the $\kappa$-$\mu$ shadowed fading model, recently  derived in \cite{paris}, is an attractive proposition. In addition to offering an excellent fit to the fading observed in a range of real-world applications (e.g. device-to-device, and body-centric fading channels \cite{Cotton}), the $\kappa$-$\mu$ shadowed fading encompasses several RF channel models such as  Nakagami-$m$, Rayleigh, Rice, $\kappa$-$\mu$ and shadowed Rician fading distributions. This new channel fading model offers far better and much	more flexible representations of practical fading LOS (line of sight), NLOS (non-LOS), and shadowed channels than the Rayleigh and Nakagami-$m$ distributions.
	Under the assumption of AF relaying and taking into account  the effect
of  pointing errors while considering both heterodyne and intensity modulation/direct (IM/DD) detection techniques,   we derive closed-form expressions
for the ergodic capacity and outage probability of dual-hop
FSO/RF systems over  M\'alaga-$\mathcal{M}$/$\kappa$-$\mu$ shadowed channels. We further pursue high signal-to-noise
ratio (SNR) analysis
to derive the diversity order.
	\vspace{-0.1cm}
 \section{Channel and System Models}
We consider a relay-assisted mixed FSO/RF transmission composed of both M\'alaga-$\mathcal{M}$ with pointing errors and  $\kappa$-$\mu$ shadowed fading environments. The source communicates with the destination through an intermediate relay, able to activate both heterodyne and IM/DD detection techniques at the reception of the optical beam.

The FSO ($S$-$R$) link irradiance is assumed to follow a M\'alaga-$\mathcal{M}$ distribution with pointing errors impairments for which the PDF of the irradiance, $I$, is given by  \cite[Eq. (5)]{ansari}\vspace{-0.15cm}
  \begin{equation}
  \label{eq:2}
f_I(x)=\frac{\!\xi^2A}{x\Gamma (\alpha)}\sum_{k=1}^{\beta}\frac{b_k}{\Gamma(k)} {\rm G}_{1,3}^{3,0} \Biggl[\frac{\alpha\beta}{g \beta+\Omega} \frac{x}{A_0}\Bigg\vert \  { \xi^2\!+1 \atop \xi^2,\alpha,k}\Biggr],
  \end{equation}
where  $\xi$ is the ratio between the equivalent beam radius and the pointing error displacement standard deviation (i.e., jitter) at the relay (for negligible pointing errors $\xi \rightarrow \infty$), $A_0$ defines the pointing loss \cite{Yang2},
$ A={ \alpha^{\frac{\alpha}{2}}\left[  {g \beta }/({g \beta +\Omega})\right] ^{\beta +\frac{\alpha}{2}}}{g^{-1-\frac{\alpha}{2}}}$ and $ {b_k}\!\!=\!\binom{\beta\!-\!1}{k\!-\!1}\!{\left(g \beta\!+\Omega \right)^{1-\frac{k}{2}}}\left[ {(g \beta +\!\Omega)}/{\alpha\beta}\right] ^{\frac{\alpha+k}{2}}\left( {\Omega}/{g}\!\right)^{k-1}\left({\alpha}/{\beta}\!\right)^{\frac{k}{2}}$, where $\alpha$, $\beta$, $g$,  and $\Omega$ are the fading parameters related to the atmospheric turbulence conditions  \cite{ansari}. Moreover in (\ref{eq:2}), ${\rm G}_{p, q}^{m, n} [\cdot]$ and $\Gamma(\cdot)$ stand for the Meijer-G  \cite[Eq. (9.301)]{grad} and the incomplete gamma \cite[Eq. (8.310.1)]{grad} functions, respectively. It is worth highlighting that the $\mathcal{M}$ distribution unifies most of the proposed statistical models characterizing  the optical irradiance in  homogeneous and isotropic turbulence \cite{ansari}. Hence both $\mathcal{G}$-$\mathcal{G}$ and ${\cal K}$ models are special cases of the M\'alaga-$\mathcal{M}$ distribution, as they mathematically derive from (\ref{eq:2})  by setting ($g=0$, $\Omega=1$) and ($g\neq0$, $\Omega=0$ or $\beta=1$), respectively \cite{ansari}.

The RF ($R$-$D$) link, experiences the  $\kappa$-$\mu$ shadowed fading with non-negative real shape parameters $\kappa$, $\mu$ and $m$,  for which the PDF of instantaneous SNR, $\gamma_2$,  is given by \cite[Eq.(4)]{paris}\vspace{-0.1cm}
\begin{eqnarray}
\label{eq:5}
f_{\gamma_2}(x)&=&\frac{\mu^\mu m^m(1+\kappa)^{\mu}}{\Gamma(\mu) \bar{\gamma}_2 (\mu \kappa+m)^m}\left(\frac{x}{\bar{\gamma}_2} \right)^{\mu-1}e^{-\frac{\mu(1+\kappa) x}{\bar{\gamma}_2}}\nonumber\\
&&\times\quad{}_1 F_1\left(m,\mu;\frac{\mu^2\kappa (1+\kappa)}{\mu \kappa+m}\frac{x}{\bar{\gamma}_2}\right),
\end{eqnarray}
where ${}_1 F_1(\cdot)$ is the confluent hypergeometric function \cite[Eq.(9.210.1)]{grad} and $\bar{\gamma}_2=\Exp[\gamma_2]$. This fading  model  jointly includes
 large-scale and small-scale propagation effects, by considering
 that only the dominant components (DSCs) are affected by Nakagami-$m$ distributed  shadowing \cite{paris}. The shadowed $\kappa$-$\mu$ distribution is an extremely versatile fading model that    includes as special cases  nearly all linear fading models pertaining to LOS and NLOS scenarios,  such as $\kappa$-$\mu$ ($m\rightarrow \infty$),  Nakagami-$m$ ($\mu=m$ and $\kappa\rightarrow 0$), Rayleigh ($\mu=m=1$ and  $\kappa\rightarrow 0 $ ),  and Rice ($\mu=1, \kappa=K$ and $m\rightarrow\infty$),  to name a few  \cite[Table I]{paris}.

Assuming AF relaying with channel state information (CSI), then the end-to end SNR can be expressed as\cite[Eq.(7)]{zedini}\vspace{-0.1cm}
\begin{equation}
\label{eq:1}
\gamma=\frac{\gamma_1 \gamma_2}{\gamma_1+\gamma_2+1},
\end{equation}
where $\gamma_{1}=(A_0h(g+\Omega))^{-r} \mu_rI^r$ is the instantaneous SNR of the FSO link
($S{ - }R$) with $r$ being the parameter that describes the detection technique at the relay (i.e., $r = 1$
is associated with heterodyne detection and $r = 2$ is associated
with IM/DD) and, $\mu_r$ refers to the electrical SNR of the FSO
hop \cite{ansari} and  $h=\xi^2/(\xi^2+1)$. In particular, for $r = 1$,   $\mu_1=\mu_{\text{heterodyne}}=\Exp[\gamma_1]=\bar{\gamma}_1$ and for $r=2$, $\mu_2=\mu_{\text{IM/DD}}=\mu_{1} \alpha\xi^2(\xi^2+1)^{-2}(\xi^2+2)(g+\Omega)/((\alpha+1)[2g(g+2\Omega)+\Omega^2(1+1/\beta)])$ \cite[Eq.(8\label{key})]{ansari}.
\section{Exact Performance Analysis}
In this section, a new mathematical framework investigating the average capacity and the outage probability of the mixed FSO/RF transmission composed of both M\'alaga-$\mathcal{M}$ with pointing errors and shadowed
$\kappa$-$\mu$ fading environments and accounting for both detection techniques is presented. To the best of the author's knowledge, there are few works that consider these metrics  of mixed FSO/RF systems, yet mostly  considering the mixed  $\mathcal{G}$-$\mathcal{G}$/Nakagami-$m$ fading (\!\!\cite{emmna},\cite{zedini} and references therein). This paper completes and extends the efforts of \cite{emmna}-\cite{zedini}  by unifying the ergodic capacity and the outage probability analysis for any turbulence/fading model under both types of detection techniques.\vspace{-0.3cm}
\subsection{Ergodic Capacity}
 Hereafter, we provide capacity formulas for the considered system by using the complementary moment
 generation function CMGF-based approach \cite{imen2} as
 \begin{equation}
 \label{eq:6}
 C\!\overset{\Delta}{=}\!\frac{ \Exp [\ln (1+\gamma)]}{2\!\ln (2)}=\frac{1}{2\!\ln (2)} \!\!\int_{0}^{\infty}\!\!\!se^{-s}M^{(c)}_{\gamma_1}(s)\!M^{(c)}_{\gamma_2}(s)ds,
 \end{equation}	
 where $M^{(c)}_X(s)=\int_{0}^{\infty}e^{-sx}F^{(c)}_X(x) dx$ stands for the CMGF with  $F^{(c)}_X(x)$  denoting the complementary cumulative distribution function (CCDF) of $X$.

 The ergodic capacity of mixed M\'alaga-$\mathcal{M}$/$\kappa$-$\mu$ shadowed fading  FSO transmission system under heterodyne and IM/DD detection techniques with pointing errors taken into account is given for \vspace{-0.1cm}
\begin{itemize}
\item \textit{Integer $m$, $\mu$, with $m \geqslant \mu$ } as\vspace{-0.2cm}
\begin{equation}
\label{eq:12}
\!\!\!C=\frac{\xi^2 A r {\mu_r}{B^{-r}}}{2 \ln(2)\Gamma(\alpha)}\sum_{k=1}^{\beta}\frac{b_k}{\Gamma(k)} \sum_{l=1}^{m}\frac{\chi_l}{\Gamma(m)} T(\theta_2,l,m),
\end{equation}
\end{itemize}
where  $B={\alpha\beta h (g+\Omega)}/[{(g\beta+\Omega)}]$, and,\vspace{-0.2cm}
\begin{equation*}
\centering
\!\!\!\!\!\!\!\!\!\!\text{}\chi_l = \left\{
\begin{array}{l l}
\!\!\!\!{\binom{m}{l}\theta_2^l} -{\binom{m-\mu}{l}\theta_1^l} & \text{ for }1\leqslant l\leqslant m-\mu, \\
\!\!\!\!{\binom{m}{l}\theta_2^l} & \text{ for }l> m-\mu,\\
\end{array} \right.
\end{equation*}
 with $\theta_1=\frac{\bar{\gamma_2}}{\mu(1+\kappa)}$ , and, $\theta_2=\frac{\bar{\gamma_2}(\mu\kappa+m)}{\mu m (1+\kappa)}$. Moreover in (\ref{eq:12}),
\begin{equation}
\label{eq:T}
T(x,y,z)={\rm H}_{10,43,11}^{01,14,11}\Biggl[\!\! \frac{\mu_r}{B^{r}}; x\Bigg\vert\ \!\!{\!(\!-\!y,\!1,\!1)\atop-\!}\!\ \!\Bigg\vert\!\ \!\!{(\sigma,\Sigma)\atop (\phi,\Phi)}\!\Bigg\vert \ \!\!\!{(\!1\!-\!z,\!1\!)\atop(\!0,1\!)}\!
\Biggr],
\end{equation}
where $\rm H[\cdot,\cdot]$ denotes the Fox-H function (FHF) of two variables \cite[Eq.(1.1)]{mittal} also known as the bivariate FHF whose Mathematica implementation may be found in \cite[Table I]{Lei}, whereby $(\sigma,\Sigma)= (1\!-\!r,r),(1\!-\!\xi^2\!-\!r,r),(1\!-\!\alpha\!-\!r,r),(1\!-\!k\!-\!r,r)$ and $(\phi,\Phi)=(0,1),(-\xi^2-r,r),(-r,r)$.  Moreover, it becomes  for \vspace{-0.1cm}
\begin{itemize}
\item \textit{Integer $m$, $\mu$, with $m  < \mu$} as\vspace{-0.2cm}
\begin{eqnarray}
\label{eq:14}
C&\!\!\!\!\!\!\!\!\!\!=\!\!\!\!\!\!\!\!\!\!&\frac{\xi^2 A r {\mu_r}}{2 \ln(2)\Gamma(\alpha){B^r}}\sum_{k=1}^{\beta}\frac{b_k}{\Gamma(k)}\!\underset{(p,q)\neq (0,0)}{\sum_{p=0}^{m}\sum_{q=0}^{\mu-m}}\binom{m}{p}\binom{\!\mu\!-\!m\!}{q}\nonumber\\
&&\!\!\!\!\!\!\!\!\!\!\theta_2^p\theta_1^q\Biggr(\sum_{i=1}^{\mu-m}\!\!\frac{\!\Delta_{1i}}{\Gamma(\!\mu\!-\!m\!-\!i\!+\!1)}T(\theta_1,q+p,\mu\!-\!m\!-\!i\!+\!1)\nonumber\\
&&\!\!\!\!\!\!\!\!\!\!-\sum_{i=1}^{m}\frac{\Delta_{2i}}{\Gamma(m-i+1)}T(\theta_2,q+p,m-i+1)\Biggl),
\end{eqnarray}
\end{itemize}
where $\Delta_{1i}=(-1)^{m} \binom{m+i-2}{i-1}\left(\frac{m}{\mu \kappa+m}\right)^m\!\!\left(\frac{\mu \kappa}{\mu \kappa+m}\right)^{-\!m\!-\!i\!+\!1}$ and $\Delta_{2i}=(-1)^{i-1} \binom{\mu-m+i-2}{i-1}\left(\frac{m}{\mu \kappa+m}\right)^{i\!-\!1}\!\!\left(\frac{\mu \kappa}{\mu \kappa+m}\right)^{\!m\!-\!\mu\!-\!i\!+\!1}$.
 \begin{IEEEproof}
 	Capitalizing on (\ref{eq:6}) and recalling the fact that the FSO link's CMGF $M^{(c)}_{\gamma_1}(s)=\mathcal{L}(F^{(c)}_{\gamma_1}(x))$ where $\mathcal{L}$ denotes the Laplace transform operator and the FSO link's CCDF is obtained as \vspace{-0.3cm}
 	 \begin{eqnarray}
\label{eq:9}
F^{(c)}_{\gamma_1}(x)\!\!\!\!&\!\!=\!\!&\!\!\!\!F^{(c)}_{I}\left( A_0 h (g+\Omega) \left( \frac{x}{\mu_r}\right)^{\frac{1}{r}}\right)  \nonumber\\
\!\!\!\!&\overset{(a)}{=}&\!\!\!\!\frac{\xi^2A}{\Gamma(\alpha)}\sum_{k=1}^{\beta}\frac{b_k}{\Gamma(k)} {\rm G}_{2,4}^{4,0} \Biggl[\!B \left(\frac{x}{\mu_r}\right)^{\frac{1}{r}}\!\! \Bigg\vert \!\! \ { \xi^2+1,1 \atop 0,\xi^2,\alpha,k}\!\Biggr].
\end{eqnarray}
where $(a)$ follows from  integrating (\ref{eq:2})   using \cite{grad}. Then, expressing the Meijer-G function in (\ref{eq:9}) in terms of Fox-H function by means of \cite[Eq.(1.111)]{mathai} and resorting to \cite[Eq.(2.19)]{mathai} with some additional manipulations using \cite[Eqs. (1.58), (1.59), \text{and} (1.60)]{mathai} yield\vspace{-0.2cm}
\begin{equation}
\label{eq:7}
M^{(c)}_{\gamma_1}(s)=\frac{\xi^2A r  \mu_r}{\Gamma(\alpha)B^{r}}\sum_{k=1}^{\beta}\frac{b_k}{\Gamma(k)}  {\rm H}_{4,3}^{1,4} \Biggl[\frac{\mu_r}{B^r} s\Bigg\vert \ { (\sigma,\Sigma) \atop (\phi,\Phi)}\Biggr],
\end{equation}
where  ${\rm H}_{p, q}^{m, n} [\cdot]$ is the Fox-H function \cite[Eq.(1.2)]{mathai}.

On the RF side, the CMGF of $\gamma_2$ under shadowed  $\kappa$-$\mu$ fading is given by \vspace{-0.3cm}
\begin{equation}
\label{eq:11}
M^{(c)}_{\gamma_2}(s)=\frac{1-M_{\gamma_2}(s)}{s}
\overset{(a)}{=}\frac{1-\frac{\left(\theta_1 s+1\right)^{m-\mu}}{\left(\theta_2 s+1\right)^m }}{s},
\end{equation}
where (a) follows from the recent result in \cite[Eq.(5)]{paris}. By assuming integer-valued $m$ and $\mu$, the RF link's CMGF can be rewritten after resorting to the transformation $\Gamma(\alpha)(1+z)^{-\alpha}={\rm H}_{1,1}^{1,1} [ z \vert\!\!\ {(1-\alpha,1) \atop (0,1)}]$ in \cite[Eq.(1.43)]{mathai} as\vspace{-0.3cm}
\begin{equation}
\label{eq:13}
M^{(c)}_{\gamma_2}(s)\overset{(a)}{\underset{\mu \leqslant m}=}\sum_{l=1}^{m}\frac{\chi_l s^{l-1}}{\Gamma(m)}{{\rm H}_{1,1}^{1,1}\left[\theta_2s\Bigg \vert \  {(1-m,1)\atop (0,1)}\right]},
\end{equation}
and \vspace{-0.35cm}
 		\begin{eqnarray}
 		\label{eq:15}
 		M^{(c)}_{\gamma_2}(s)&\overset{(b)}{\underset{\mu> m}=}&\underset{(p,q)\neq (0,0)}{\sum_{p=0}^{m}\sum_{q=0}^{\mu-m}}\binom{m}{p}\binom{\!\mu\!-\!m\!}{q}\theta_2^p\theta_1^qs^{p+q-1}\nonumber\\&&
 		\!\!\!\!\!\!\!\!\!\!\!\!\!\!\!\!\!\!\!\!\!\!\!\!\!\!\!\!\!\!\!\!\!\!\!\!\!\!\!\!\!\!\!\!\!\!\!\!\!\! \Biggr(\!\!\sum_{i=1}^{\mu-m}\!\!\frac{\Delta_{1i}{\rm H}_{1,1}^{1,1}\!\!\left[\!\theta_1 s \vert \ \!\! {\!(\!m\!+i\!-\mu,1)\atop (0,1)}\!\right]}{\Gamma(\mu\!-\!m-\!i+\!1)}\!-\!\sum_{i=1}^{m}\!\!\frac{\Delta_{2i}{\rm H}_{1,1}^{1,1}\!\!\left[\!\theta_2 s \vert \ \!\! {(i\!-\!m,1)\atop \!\!(0,1)}\!\right]}{\Gamma(m\!-\!i\!+\!1)}\!\!\Biggl),		
 		\end{eqnarray}
where (a) and (b) follow after applying the binomial expansion and the partial fraction decomposition \cite[Eq.(27)]{paris4}, respectively. Plugging (\ref{eq:7}), (\ref{eq:13}) and (\ref{eq:15}) into (\ref{eq:6}) and resorting to \cite[Eq. (2.2)]{mittal} complete the proof.
 \end{IEEEproof}
 \vspace{-0.2cm}
 \subsection{Outage Probability}
 The quality of service (QoS) of the considered mixed FSO/RF system  is ensured by keeping the instantaneous end-to-end SNR, $\gamma$, above a threshold $\gamma_{th}$. The probability of outage in the mixed FSO/RF  relaying setup is expressed as\vspace{-0.2cm}
 	\begin{equation}
 	\label{eq:outfix1}
 	P_{\text{out}}={\rm Pr}[\gamma<\gamma_{th}]={\rm Pr}\left[\frac{\gamma_1 \gamma_2}{\gamma_1+\gamma_2+1}<\gamma_{th}\right].
 	\end{equation}
 Marginalization over $\gamma_1$, and letting $u\!=\!1\!+\!\gamma/\gamma_{th}$ in (\ref{eq:outfix1}) yield \vspace{-0.3cm}
  \begin{equation}
  \label{out}
  \!\!P_{\text{out}}(\gamma_{th})\!=\!1-\gamma_{th}\!\int_{1}^{\infty}\!\!\!\!\!F^{(c)}_{\gamma_2}\left(\gamma_{th}\!+\!\frac{1\!+\!\gamma_{th}}{u\!-\!1}\right)f_{\gamma_1}(u\gamma_{th}) du,
  \end{equation}
  where $F^{(c)}_{\gamma_2}$  is the CCDF of $\gamma_2$ and  $f_{\gamma_1}$ is the PDF of the first-link SNR obtained from deriving (\ref{eq:9}) with respect to $x$ as\vspace{-0.2cm}
  \begin{equation}
  \label{eq:pdfFSO}
  \!\!f_{\gamma_1}(x)\!\!=\!\!\frac{\xi^2AB^r}{\Gamma(\alpha)\mu_r}\!\!\sum_{k=1}^{\beta}\!\!\frac{b_k}{\Gamma(k)}\!{\rm H}_{1,3}^{3,0}\!\Biggl[\!\frac{B^rx}{\mu_r}\Bigg\vert \!\! \ { (\!\xi^2\!+\!1\!-\!r,r\!) \atop\!(\!\xi^2\!-r\!,\!r\!)\!,(\!\alpha\!-\!r\!,\!r)\!,\!(\!k\!-\!r\!,\!r\!)}\!\!\Biggr].
  \end{equation}
  Plugging (\ref{eq:pdfFSO}) and the RF link’s CCDF expression recently derived in \cite[Eq.(10)]{paris4} for integer  $m$, $\mu$ with  $m \geqslant \mu$   into the above integral and making  a Taylor expansion of exponential and power terms, we infer that \vspace{-0.3cm}
  \begin{eqnarray}
  \label{eq:out1}
  \!\!\!\!\!P_{\text{out}}(\gamma_{th})\!\!&\!\!=\!\!&\!\!1-\frac{\xi^2AB^r\gamma_{th}e^{-\frac{\gamma_{th}}{\theta_2}}}{\Gamma(\alpha)\mu_r}\sum_{k=1}^{\beta}\frac{b_k}{\Gamma(k)}
  \!\nonumber\\
  &&\!\!\!\sum_{i=0}^{m\!-\mu}\sum_{j=0}^{m\!-\!i\!-1}\sum_{p=0}^{j}\frac{\binom{j}{p}\Upsilon_i}{j!\theta_2^{j}}\gamma_{th}^{j-p} (\gamma_{th}+1)^{p} \times \mathcal{I},
  \end{eqnarray}
  with $\Upsilon_i=\binom{m-\mu}{i}\left(\frac{m}{\mu \kappa+m}\right)^i\!\!\left(\frac{\mu \kappa}{\mu \kappa+m}\right)^{\!m\!-\!\mu\!-\!i}$, and $\mathcal{I}$ given by \vspace{-0.3cm}
  \begin{eqnarray}
  \label{eq:I}
  \mathcal{I} &=&\sum_{q=0}^{\infty}\sum_{l=0}^{\infty} \frac{(-1)^{q}(p+q)_l}{q!l!\theta_2^{q}} (\gamma_{th}+1)^{q}\int_{1}^{\infty} u^{-p-q-l} \nonumber\\
  &&\!\!\!\!\!\!\!\!\!{\rm H}_{1,3}^{3,0} \Biggl[\!\frac{B^r\gamma_{th}}{\mu_r}u\Bigg\vert \  { (\xi^2\!+\!1\!-\!r,r) \atop \!\!\! (\!\xi^2\!-\!r,r\!),(\!\alpha\!-\!r,r\!),(\!k\!-\!r,r\!)}\!\!\Biggr]du.
  \end{eqnarray}
  Substituting  (\ref{eq:1}) into (\ref{eq:out1}) after resorting to \cite[Eq. (2.54)]{mathai} yields the outage probability of mixed FSO/RF in M\'alaga-$\mathcal{M}$/$\kappa$-$\mu$ shadowed fading ($\mu \leqslant m$ ) environments  with pointing errors under both detection techniques as\vspace{-0.2cm}
\begin{equation}
\label{eq:outex}
\!P_{\text{out}}(\gamma_{th})\!=\!1\!-\!\frac{\xi^2AB^r\gamma_{th}}{e^{\frac{\gamma_{th}}{\theta_2}}\Gamma(\alpha)\mu_r}\sum_{k=1}^{\beta}\!\frac{b_k}{\Gamma(k)}\!\!\sum_{i=0}^{m\!-\!\mu}\sum_{j=0}^{m\!-\!i\!-1}\!\!\frac{\Upsilon_i}{\theta_2^{j}}\Xi(\theta_2),
\end{equation}\vspace{-0.1cm}
where \vspace{-0.2cm}
\begin{equation}
\label{eq:Xi}
\!\!\Xi(x)\!=\!\sum_{p=0}^{j}\!\sum_{q=0}^{\infty}\!\sum_{l=0}^{\infty}\!\frac{(\!-1\!)^{q}(\!p\!+\!q\!)_l\binom{j}{p}\gamma_{th}^{j-p}}{j!q!l!x^{q}(\!\gamma_{th}\!+\!1\!)^{-p-q}}{\rm H}_{2,4}^{4,0}\!\Biggl[\!\frac{B^r\!\gamma_{th}}{\!\mu_r}\!\Bigg\vert \  {\!\!\!(\!\sigma_1\!,\!\Sigma_1\!) \atop \!\!\!(\!\phi_1\!,\!\Phi_1\!)}\!\!\Biggr],
\end{equation}
with $(a)_n$ standing for the Pochhammer symbol \cite{grad}, $(\sigma_1,\Sigma_1) =(\xi^2\!+\!1\!-\!r,r),(\!l\!+\!p\!+\!q,\!1\!)$, and $(\phi_1,\Phi_1)=(l\!+\!p\!+\!q\!-\!1,\!1),(\xi^2\!-\!r,r),(\alpha\!-\!r,r),(k\!-\!r,r)$.

Similar to (\ref{eq:outex}) and using \cite[Eq.(9)]{paris4}, the outage probability of mixed FSO M\'alaga-$\mathcal{M}$/RF shadowed $\kappa$-$\mu$ ($m < \mu$) is \vspace{-0.2cm}
\begin{eqnarray}
\label{eq:out2}
\!\!\!\!\!\!\!\!P_{\text{out}}(\gamma_{th})&\!\!=\!\!&1-\frac{\xi^2AB^r\gamma_{th}}{\Gamma(\alpha)\mu_r}\sum_{k=1}^{\beta}\frac{b_k}{\Gamma(k)}\!\nonumber\\
&&\!\!\!\!\!\!\!\!\!\!\!\!\!\!\!\!\!\!\!\!\!\!\!\!\!\!\!\!\!\!\!\!\!\!\!\!\Biggr(\!e^{-\frac{\gamma_{th}}{\theta_1}}\sum_{i=1}^{\mu\!-\!m}\sum_{j=0}^{\mu\!-\!m\!-i}\!\!\frac{\Delta_{1i}}{\theta_1^{j}}\Xi(\theta_1)+e^{-\frac{\gamma_{th}}{\theta_2}}\!\sum_{i=1}^{m}\sum_{j=0}^{m\!-\!i}\frac{\Delta_{2i}}{\theta_2^{j}}\Xi(\theta_2)\!\!\Biggl).
\end{eqnarray}
\vspace{-0.45cm}
\section{Asymptotic Analysis}
To gain more insights into the effect of turbulence/fading parameters on both the ergodic capacity and the outage probability, we study hereafter their asymptotic behaviors.  To this end, we invoke  the asymptotic expansions of  the Fox-H function \cite[Theorems 1.7 and 1.11]{kilbas} and the Mellin-Barnes integrals involving the bivariate Fox-H function \cite[Eq. (2.56)]{mathai}.\vspace{-0.35cm}
\subsection{Asymptotic Ergodic Capacity}
We assume that the average SNR of the RF
link $\bar{\gamma}_2$ goes to infinity for a fixed and finite valued average
SNR in the FSO link. Then, resorting to the Mellin-Barnes representation of the bivariate FHF \cite[Eq.(2.56)]{mathai} in  (\ref{eq:12}), and evaluating the residue  at the poles $\{ -m, -1-l \}$ yield the asymptotic capacity,   when $m \geq \mu$ as \vspace{-0.4cm}
\begin{eqnarray}
\label{eq:12bis}
C^\infty\!\!\!\!&\!\!\!\!=\!\!\!\!&\!\!\!\!\frac{\xi^2 A r {\mu_r}}{2 \ln(2)\Gamma(\alpha){B^r}}\!\!\sum_{k=1}^{\beta}\sum_{l=1}^{m}\!\frac{b_k\chi_l}{\Gamma(k)} \Bigg(\frac{{\rm H}_{5,4}^{2,5} \Biggl[\!\!\frac{\mu_r}{B^r \theta_2}\Bigg\vert \ \!\!{(-l,1),(\sigma,\Sigma)\atop \!\!(m-1-l,1),(\phi,\Phi)}\!\!\Biggr]}{\theta_2^{1+l}\Gamma(m)}\nonumber \\
&& + \theta_2^{-m}{\rm H}_{5,3}^{1,5}\Biggl[\frac{\mu_r}{B^r}\!\Bigg\vert \ { (m-l,1),(\sigma,\Sigma)\atop (\phi,\Phi)}\Biggr]\Bigg).
\end{eqnarray}
It is worth noting that (\ref{eq:12bis})  is much easier and
faster to calculate than the exact capacity in (\ref{eq:12}).
Moreover, $C^\infty$ when $m<\mu$ follows in the same line of (\ref{eq:12bis}) while considering (\ref{eq:14}).

% The asymptotic behavior when $\mu_r$, $\bar{\gamma}_2\rightarrow\infty$  is inferred by computing the residue of the Fox-H function in (\ref{eq:12bis}) at $(\frac{\sigma'-1}{\Sigma'})$ using \cite[Eq.(1.5.9)]{kilbas} as \vspace{-0.2cm}
%\begin{eqnarray}
%	\label{eq:12bis1}
%	\!\!\!\!\!\!\!\!\!\!\!\!\!C^\infty&\!\!\!\!=\!\!\!\!&\frac{\xi^2 A r \theta_2^{-m}}{2 \ln(2)\Gamma(\alpha)}\sum_{k=1}^{\beta}\sum_{l=1}^{m}\sum_{t=1}^{5}\frac{b_k{\chi_l}}{\Gamma(k)}\frac{1}{\Sigma'_t}\nonumber\\
%	&&\!\!\!\!\!\!\!\!\!\!\!\!\!\!\!\!\frac{\Gamma(\frac{1-\sigma'_t}{\Sigma'_t}){\prod_{\underset{s \neq t}{s=1}}^{5}}\Gamma(1-\sigma'_s+(\sigma'_t-1)\frac{\Sigma'_s}{\Sigma'_t})}{\prod_{s=2}^{3}\Gamma(1-\phi_s+(\sigma'_t-1)\frac{\Phi_s}{\Sigma'_t})}\left(\!\frac{\mu_r}{B^r}\!\right)^{\frac{\sigma'_t\!-\!1}{\Sigma'_t}\!+\!1}.
%	\end{eqnarray}
\vspace{-0.35cm}
\subsection{Asymptotic Outage Probability}
At high SNR values, the outage probability of the mixed FSO/RF relaying system can
be expressed as $P_{\text{out}}\simeq (G_c \text{SNR} )^{-G_d}	$,  where $G_c$ and $G_d$
 denote  the
coding gain and the diversity order of the system, respectively. Hence, as $\mu_r \rightarrow \infty$ while keeping the low-order terms in (\ref{eq:outex}), i.e. $q+l < 1$, and   then  applying \cite[Eq. (1.8.5)]{kilbas}  yield  the asymptotic CDF  when $m \geq \mu$  as \vspace{-0.2cm}
\begin{eqnarray}
\label{eq:out1high}
\!\!\!\!\!\!\!P_{\text{out}}^{\infty}\!\!&\!\!\!\!=\!\!\!\!&\!\!1\!-\!\frac{\xi^2\!Ae^{-\frac{\gamma_{th}}{\theta_2}}}{\Gamma(\alpha)}\!\sum_{k=1}^{\beta}\!\frac{b_k}{\Gamma(k)}\!\sum_{i=0}^{m\!-\mu}\sum_{j=0}^{m\!-\!i\!-1}\!\!\sum_{p=0}^{j}\!\frac{\binom{j}{p}\Upsilon_i\gamma_{th}^{j-p}\theta_2^{-j}}{j!(\gamma_{th}+1)^{-p}}\nonumber\\
&&\sum_{t=1}^{4}\frac{1}{\Phi_{1t}}\frac{{\prod_{\underset{s \neq t}{s=1}}^{4}}\Gamma\left(\phi_{1s}-\phi_{1t}\frac{\Phi_{1s}}{\Phi_{1t}}\right)}{\prod_{s=1}^{2}\Gamma\left(\sigma_{1s}-\phi_{1t}\frac{\Sigma_{1s}}{\Phi_{1t}}\right)}\left(\!\!\frac{B^r\gamma_{th}}{\mu_r}\!\!\right)^{\frac{\phi_{1t}}{\Phi_{1t}}+1}.
\end{eqnarray}
%\begin{eqnarray}
%\label{eq:out1high}
%\!\!\!\!\!\!\!P_{\text{out}}^{\infty}\!\!&\!\!\!\!=\!\!\!\!&\!\!1\!-\!\frac{\xi^2\!Ae^{-\frac{\gamma_{th}}{\theta_2}}}{\Gamma(\!\alpha\!)}\!\sum_{k=1}^{\beta}\!\frac{b_k}{\Gamma(\!k\!)}\!\sum_{i=0}^{m\!-\mu}\sum_{j=0}^{m\!-\!i\!-1}\!\!\sum_{p=0}^{j}\!\frac{\binom{j}{p}\Upsilon_i\gamma_{th}^{j-p}\theta_2^{-j}}{j!(\gamma_{th}+1)^{-p}}\nonumber\\
%&&\sum_{n=1}^{4}\tau_n\left(\frac{B^r\gamma_{th}}{\mu_r}\right)^{\Pi_n},
%\end{eqnarray}
 Compared
to (\ref{eq:outex}) which is expressed in terms of Fox-H
function, (\ref{eq:out1high}) includes only finite summations of elementary
functions.  The diversity gain of the studied system over atmospheric turbulence conditions is inferred after applying $e^{-\frac{\gamma_{th}}{\theta_2}}\underset{\bar{\gamma}_2\gg 1}{\approx}1-\frac{\gamma_{th}}{\theta_2}$ to  (\ref{eq:out1high}) as $G_d=\min\left\{\mu, \frac{\xi^2}{r},\frac{\alpha}{r},\frac{\beta}{r}\right\}$.

For $\mu=m$, $\kappa\rightarrow0$, $g=0$ and $\Omega=1$, the CDF in  (\ref{eq:out1high}) reduces to $P_{\text{out}}^{\infty}$ for  $\mathcal{G}$-$\mathcal{G}$/Nakagami-$m$ fading
 channels as\vspace{-0.2cm}
\begin{eqnarray}
\label{eq:out1high2}
\!\!\!\!\!\!\!P_{\text{out}}^{\infty}\!\!&\!\!\!\!=\!\!\!\!&\!\!1\!-\!\frac{\xi^2e^{-\frac{m\gamma_{th}}{\bar{\gamma}_2}}}{\Gamma(\alpha)\Gamma(\beta)}\sum_{j=0}^{m\!-1}\sum_{p=0}^{j}\sum_{t=1}^{4}\!\frac{\binom{j}{p}m^j\gamma_{th}^{-p}(\alpha\beta h)^{r(\frac{\phi_{1t}}{\Phi_{1t}}+1)}}{j!\Phi_{1t}(\gamma_{th}+1)^{-p}}\nonumber\\
&&\frac{{\prod_{\underset{s \neq t}{s=1}}^{4}}\Gamma\left(\phi_{1s}-\phi_{1t}\frac{\Phi_{1s}}{\Phi_{1t}}\right)}{\prod_{s=1}^{2}\Gamma\left(\sigma_{1s}-\phi_{1t}\frac{\Sigma_{1s}}{\Phi_{1t}}\right)}\left(\!\!\frac{\gamma_{th}}{\mu_r}\!\!\right)^{\frac{\phi_{1t}}{\Phi_{1t}}+1}\left(\frac{\gamma_{th}}{\bar{\gamma}_2}\right)^{j},
\end{eqnarray}
thereby inferring \cite[Eq.(29)]{zedini}, i.e., $G_d=\min\left\{m, \frac{\xi^2}{r},\frac{\alpha}{r},\frac{\beta}{r}\right\}$.
Similar to (\ref{eq:out1high}) while considering (\ref{eq:out2}), the asymptotic outage probability can be derived  in
closed form when $m< \mu$. However, the derived expression is omitted due to
space limitations.

%, and $\tau_1=\frac{\Gamma(\xi^2-pr)\Gamma(\alpha-pr)\Gamma(\beta-pr)}{\Gamma(\xi^2+1-pr)}$, and $\tau_2=\frac{\Gamma(p-\frac{\xi^2}{r})\Gamma(\alpha-\xi^2)\Gamma(\beta-\xi^2)}{r\Gamma(p+1-\frac{\xi^2}{r})}$, and $\tau_3=\frac{\Gamma(p-\frac{\alpha}{r})\Gamma(\xi^2-\alpha)\Gamma(\xi^2-\beta)}{r\Gamma(p+1-\frac{\alpha}{r})\Gamma(\xi^2+1-\alpha)}$ and $\tau_4=\frac{\Gamma(p-\frac{\beta}{r})\Gamma(\xi^2-\beta)\Gamma(\alpha-\beta)}{r\Gamma(p+1-\frac{\beta}{r})\Gamma(\xi^2+1-\beta)}$.
% Furthermore, the diversity gain of mixed FSO/RF AF relay systems over Gamma-Gamma/Rayleigh fading follows from (\ref{eq:out1high2}) to be equal to $G_d=\min\{1,\frac{\xi^2}{r},\frac{\alpha}{r},\frac{\beta}{r}\}$. The same result is also obtained in \cite[Eq.(29)]{zedini}.
\vspace{-0.4cm}
\section{Numerical Results}

\begin{figure}
 		\centering
 		\hspace{-0.8cm}
 		\subfigure[]{
 			\includegraphics[width=5.4cm] {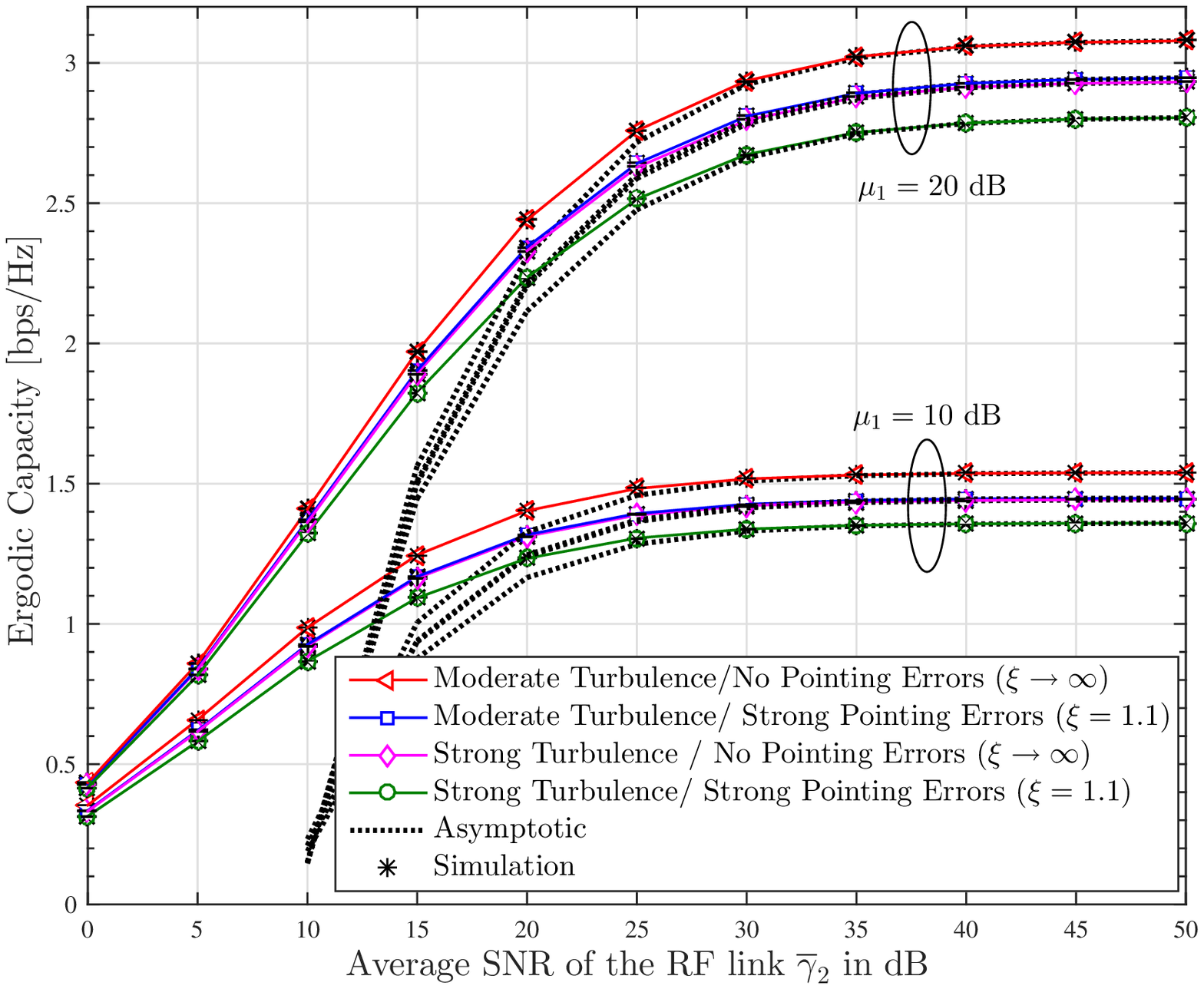}
 			\label{fig:fig1}
 		}
 		\hspace{0.1cm}
 		\subfigure[]{\hspace{-0.8cm}
 			\includegraphics[width=5.4cm] {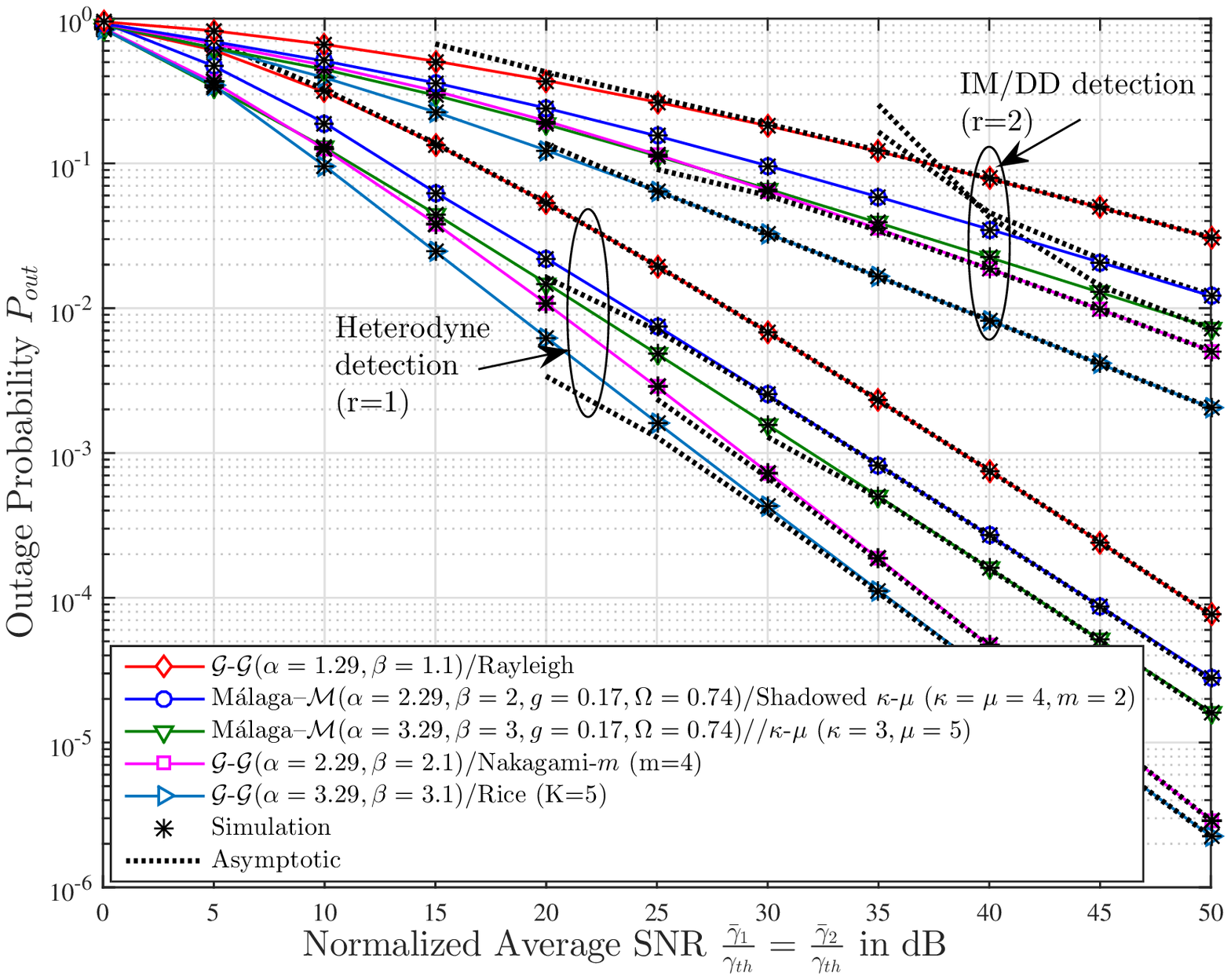}
 			\label{fig:fig2}}\vspace{-0.2cm}
 		\caption{Performance of relay-assisted mixed M\'alaga/$\kappa$-$\mu$ shadowed fading.}
 		\hspace{0.1cm}

 	\end{figure}

Fig.\ref{fig:fig1} investigates the impacts of the turbulence-induced fading and pointing errors on the system performance when the RF link is subject to Rician shadowed fading distribution ($\kappa=5$, $\mu=1$, $m=2$). As expected, the ergodic capacity deteriorates by decreasing the pointing error displacement standard deviation, i.e., for smaller $\xi$, or decreasing the turbulence fading parameter, i.e., smaller $\alpha$ and $\beta$, where we associate the strong turbulence to $(\alpha,\beta)=(2.29,2)$ and the moderate turbulence to $(\alpha,\beta)=(4.2,3)$. At high SNR, the asymptotic expansion in (\ref{eq:12bis}) matches very well its exact counterpart, which confirms the validity of our mathematical analysis for different parameter settings.

Fig.\ref{fig:fig2} depicts the outage probability of mixed FSO/RF relay systems in M\'alaga-$\mathcal{M}$ and shadowed $\kappa$-$\mu$ fading channels for both heterodyne and IM/DD detection at the relay.  Throughout our
numerical experiments, we found out that regardless of the average SNRs and turbulence/fading settings, accurate analytical curves can be obtained by truncating the infinite sums at $q=10$ and $l=5$ terms. In the legend, please note that we have identified some particular turbulence and fading distribution cases that simply stem from the general M\'alaga amd $\kappa$-$\mu$ shadowed fading scenarios, respectively. The exact match with Monte-Carlo simulation results confirms the precision of the theoretical analysis of section III.B.  Moreover, we notice that the exact and asymptotic expansion in  (\ref{eq:out1high}) agree very well at high SNRs.
 \vspace{-0.3cm}
 \section{Conclusion}
We have presented a unified analytical framework for relay-assisted mixed FSO/RF systems that remarkably accommodates generic turbulence/fading  models including M\'alaga-$\mathcal{M}$ with pointing errors and  shadowed $\kappa$-$\mu$  distribution that account for shadowed LOS and NLOS scenarios. The results demonstrate the unification of various FSO turbulent/RF fading scenarios into a single closed-form expression  for the ergodic capacity and the outage probability while accounting for both IM/DD and heterodyne detection techniques at the relay. 
\end{document}